# The infrared imaging spectrograph (IRIS) for TMT: sensitivities and simulations

Shelley A. Wright[a*], Elizabeth J. Barton[b], James E. Larkin[c], Anna M. Moore[d], David Crampton[e,f], Luc Simard[e,f], and IRIS team[f]

[a]University of California, Berkeley, 601 Campbell Hall, Berkeley, CA, 94720, USA;
[b]University of California, Irvine, 4129 Frederick Hall, Irvine, CA, 92697, USA;
[c]University of California, Los Angeles, 430 Portola Plaza, Los Angeles, CA, 90095, USA;
[d]California Institute of Technology, 1200 E. California Rd., Pasadena CA, 91125, USA;
[e]Herzberg Institute of Astrophysics, 5071 W. Saanich Rd, Victoria, B.C., V9E 2E7, Canada;
[f]Thirty Meter Telescope Project, 2632 E. Washington Blvd, Pasadena, CA, USA 91107

**Abstract**
We present sensitivity estimates for point and resolved astronomical sources for the current design of the InfraRed Imaging Spectrograph (IRIS) on the future Thirty Meter Telescope (TMT). IRIS, with TMT's adaptive optics system, will achieve unprecedented point source sensitivities in the near-infrared (0.84 – 2.45 μm) when compared to systems on current 8-10m ground based telescopes. The IRIS imager, in 5 hours of total integration, will be able to perform a few percent photometry on 26 - 29 magnitude (AB) point sources in the near-infrared broadband filters (Z, Y, J, H, K). The integral field spectrograph, with a range of scales and filters, will achieve good signal-to-noise on 22 - 26 magnitude (AB) point sources with a spectral resolution of R=4,000 in 5 hours of total integration time. We also present simulated 3D IRIS data of resolved high-redshift star forming galaxies (1 < z < 5), illustrating the extraordinary potential of this instrument to probe the dynamics, assembly, and chemical abundances of galaxies in the early universe. With its finest spatial scales, IRIS will be able to study luminous, massive, high-redshift star forming galaxies (star formation rates ~ 10 – 100 M⊙ yr$^{-1}$) at ~100 pc resolution. Utilizing the coarsest spatial scales, IRIS will be able to observe fainter, less massive high-redshift galaxies, with integrated star formation rates less than 1 M⊙ yr$^{-1}$, yielding a factor of 3 to 10 gain in sensitivity compared to current integral field spectrographs. The combination of both fine and coarse spatial scales with the diffraction-limit of the TMT will significantly advance our understanding of early galaxy formation processes and their subsequent evolution into present-day galaxies.

## 1. INTRODUCTION

In the last decade there have been considerable advances in near-infrared astronomical instrumentation designed for adaptive optic systems. Examples are the latest integral field spectrographs that have been developed for sampling the diffraction-limit of 8-10m class telescopes, such as OSIRIS[1] for Keck Observatory, SINFONI[2] for Very Large Telescope, and NIFS[3] for Gemini-North. These innovative spectrographs have made significant achievements and are stimulating the design of future instrumentation projects for the James Webb Space Telescope (JWST) and Extremely Large Telescopes.

IRIS (InfraRed Imaging Spectrograph)[4] is currently being designed as a first light instrument to take advantage of the diffraction-limit of the future optical and infrared Thirty Meter Telescope[5] (TMT) on Mauna Kea, Hawaii. IRIS will house a near-infrared (0.84 - 2.45 μm) imaging camera[6] (0.004" per pixel) and integral field spectrograph (0.004" - 0.05" per element) to sample the high angular resolution achieved with TMT's narrow field infrared adaptive optics system (NFIRAOS[7]). The integral field spectrograph uses a lenslet array for the finest spatial scales (0.004", 0.009") and a mirror-slicer for the coarsest spatial scales (0.025", 0.05")[8].

---
[*] Send Correspondence to saw@astro.berkeley.edu

TMT will offer advantages in sensitivity and angular resolution compared to existing ground-based observatories and space-based telescopes. The effective collecting area of TMT's primary will be 8 to 10 times larger than current 8-10m ground-based telescopes. The angular resolution achieved using diffraction-limited techniques will be 3 to 5 times greater than current 8-10m ground-based and future space-based facilities. Each of these gains will allow unprecedented advances in a broad range of astronomical research ranging from the solar system, extrasolar planets, stellar evolution, star formation, the Galactic Center, nearby galaxies, high-redshift galaxies ($1 < z < 5$; age of the universe from 6 to 1 Gyr), and first-light galaxies ($z < 8$; age of universe less than 600 Myr). These science cases[9] have been individually investigated and have directly influenced the design requirements for IRIS.

In this paper, we present point source sensitivities for both the imager and the four spatial scales of the integral field spectrograph of IRIS with NFIRAOS on TMT. We discuss input parameters for our simulations in Section 2. Point source sensitivities are summarized for the imager in Section 3 and for the spectrograph in Section 4. The parameter space to completely explore resolved sources (e.g., debris disks, nearby galaxies, solar system objects, and high-redshift galaxies) is very large and is beyond the scope of this paper. Instead, to illustrate the predicted sensitivities on resolved sources, we have focused on example observations of high-redshift ($1 < z < 5$) star forming galaxies using the IRIS integral field spectrograph in Section 5. We also briefly discuss the scientific gains achievable with IRIS observations of galaxies in the very early universe.

## 2. SIGNAL-TO-NOISE RATIOS & SIMULATOR

Sensitivity calculations have been performed for the imager (0.004") and spectrograph in all four spatial samplings (0.004", 0.009", 0.025", and 0.05") for point and resolved astronomical sources. For each combination of waveband and spatial sampling, the sensitivity calculator assumes an integrated background, filter profile, total integrated throughput, point-spread functions for NFIRAOS as observed from Mauna Kea, detector readnoise and dark current, spectral resolution, blaze function for a grating, and noise generated from the spectral extraction process within the spectrograph data reduction pipeline. Table 1 summarizes the primary requirements assumed for IRIS in the sensitivity calculator and simulator. Throughout this paper, we investigate sensitivities in Y ($\lambda_{cen} = 1.09$ μm), J ($\lambda_{cen} = 1.27$ μm), H ($\lambda_{cen} = 1.63$ μm), and K ($\lambda_{cen} = 2.18$ μm) broad bands.

| Parameters Used in Simulator | Value |
|---|---|
| Wavelength Range | $0.84 \leq \lambda \leq 2.45$ μm in bands *Z, Y, J, H, K* |
| Effective Collecting Area of TMT | 630 m$^2$ |
| Hawaii-4RG Readnoise | 5e- |
| Hawaii-4RG Darkcurrent | 0.006 e-/s |
| TMT+NFIRAOS+IRIS Imager Throughput | 42 – 48 % |
| TMT+NFIRAOS+IRIS Spectrograph Throughput | 32 – 42 % |
| Spectral Resolution (R) | 4000 |

**Table 1**: Major parameters used in all sensitivity calculations and simulations throughout this paper. Total throughput of the system is dependent on wavelength for the imager and spectrographs. We give the full range of throughputs in the table, but use the wavelength specific throughput dependent on the filter in the simulator.

In Figure 1, we present the infrared atmospheric throughput as observed from Mauna Kea, with the broadband filter transmission curves for Z, Y, J , H, and K and their respective grating transmission. The sky background is modeled as a combination of sky continuum from Rayleigh scattering in the atmosphere and blackbody emission with a given temperature and emissivity for the atmosphere (T=258K, $\varepsilon$=0.1), the zodiacal light (T=5800K, $\varepsilon=3\times10^{-14}$), the telescope (T=73K, $\varepsilon$=0.02), and NFIRAOS with IRIS (T=243K, $\varepsilon$=0.01). We also include empirically measured OH emission lines throughout IRIS wavelength coverage (courtesy of Gemini Observatory). It is important to note that Z and Y broadband sky backgrounds are influenced more by the lunar phase and elevation of the observations (due to the zodiacal light) than J, H, and K broadbands. For all sensitivity calculations, we assume the observations are taken during a "dark" lunar phase and away from the ecliptic pole.

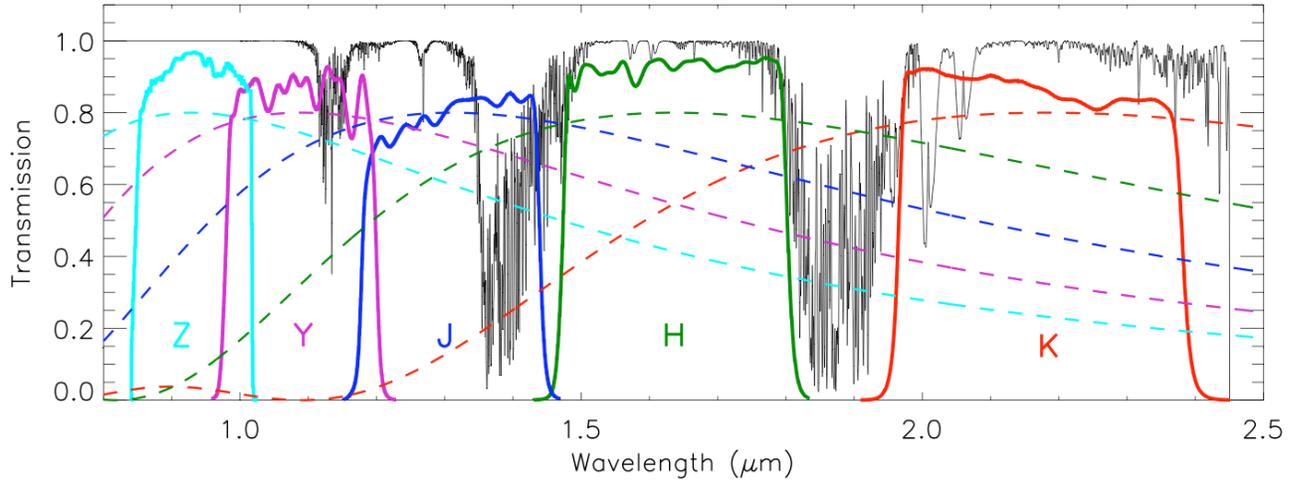

**Figure 1**: Infrared atmospheric throughput (solid black) from Mauna Kea observed at an airmass of 1.0 and water vapor of 1.6mm (courtesy of Gemini Observatory). Filter transmission curves for the broadband filters for Z (solid teal), Y (solid purple), J (solid blue), H (solid green), and K (solid red). For the same respective colors, the grating blaze functions are shown assuming first order (m=1) for each grating and 80% peak efficiency.

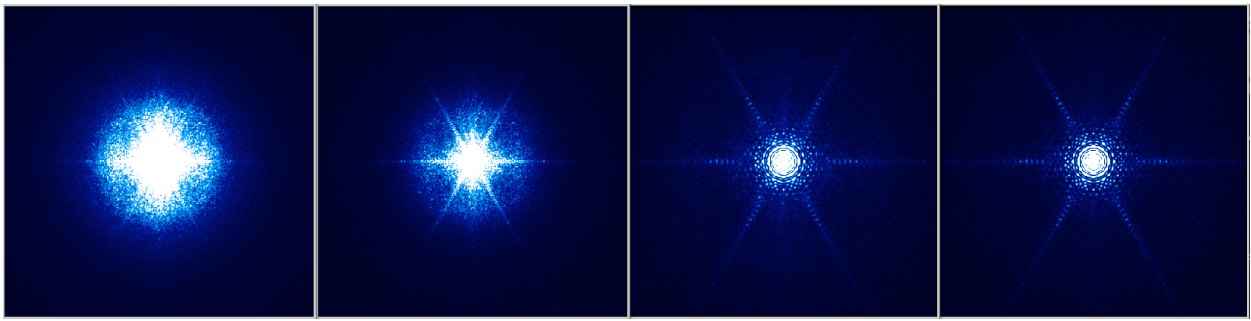

**Figure 2**: On-axis Point Spread Functions for 0.88, 1.25, 1.65, 2.20 µm (left to right) in the 0.004" spatial scale. Strehl ratios for these wavelengths are 43%, 70%, 81%, and 89%. All PSFs have the same linear image stretch in this figure.

The NFIRAOS team has generated predicted point-spread functions (PSFs) from the AO system on Mauna Kea for four wavelengths (0.88, 1.25, 1.65, and 2.2 µm), with a spatial sampling of 0.004" or less. In Figure 2, we show images of on-axis PSFs at each of these wavelengths with the same linear image stretch. The Strehl ratios achieved in these images are 43%, 70%, 81%, and 89% (left to right, 0.88 to 2.2 µm). The sensitivity calculator for the imager and spectrograph use the predicted PSFs resampled to the desired spatial scales and convolved with assumed intrinsic flux distributions to derive signal-to-noise ratios (SNR) for both point and resolved sources in each broadband observation (Y, J, H, and K).

## 3. IMAGER POINT SOURCE SENSITIVITIES

The imager is currently designed to use a Hawaii-4RG detector (see Table 1 for expected noise) with 0.004" spatial scale, yielding a 17"x17" field-of-view. The imager specifications and design have been described elsewhere by our team[6]. Figure 3 shows the predicted signal-to-noise ratios in 5 hours of total integration for four broadband observations. Imager sensitivity calculations assume an integrated background that includes OH sky emission lines, sky continuum, and sky background (as described in Section 2). These calculations assume an aperture size for each band of $2\lambda_{cen}/D$, where $\lambda_{cen}$ is the central wavelength of the filter. We caution that these calculations serve only as an approximation since we average over a range of atmospheric conditions (e.g., thermal background, water vapor content) and instrumental performance (e.g., Strehl ratio). Signal-to-noise ratio of 100 is achieved down to Y = 28.6 mag, J = 28.2 mag, H = 27.3 mag, and K = 27.0 mag. At longer wavelengths ( > 2 μm in K broadband), even though the Strehl ratio is greater, the thermal background from the telescope and atmosphere is higher and therefore lowers the overall signal-to-noise. In H (1.6 μm) broadband, the dominant integrated background comes from OH-sky emission lines and therefore yields a lower signal-to-noise performance in the imager compared to both Y and J broadbands.

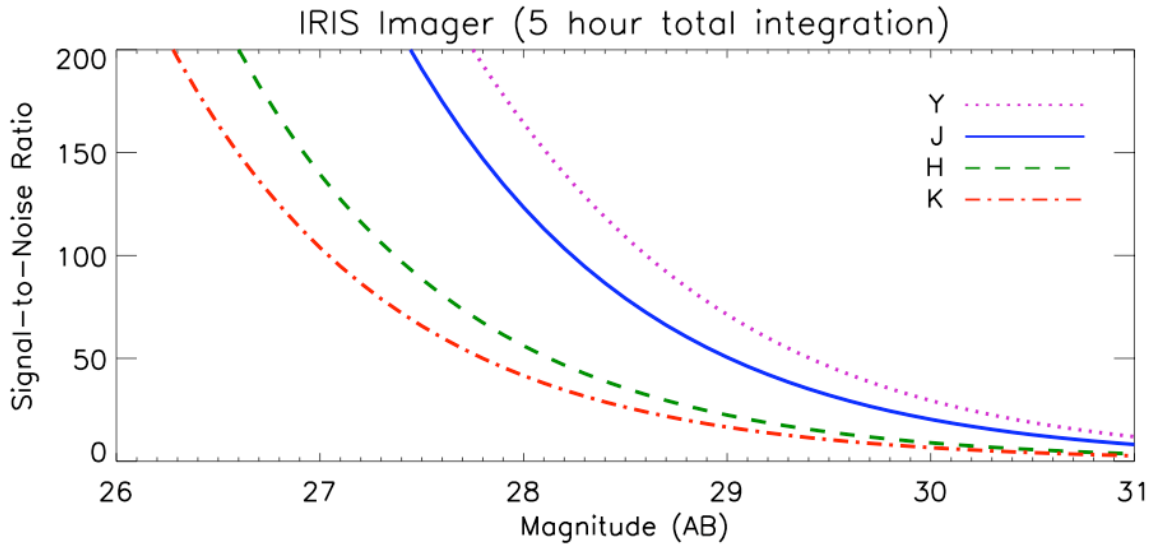

**Figure 3**: Estimated signal-to-noise ratio verse a given magnitude (AB) illustrating sensitivities for the imager (0.004" per pixel) in each broadband filter (Y, J, H, K) using a $2\lambda_{cen}/D$ aperture size for a single point source. A total integration time of 5 hours was composed of single exposures of 900 seconds stacked 20 times in Y, J, and H, and K.

## 4. SPECTROGRAPH POINT SOURCE SENSITIVITIES

The integral field spectrograph will use a Hawaii-4RG detector (see Table 1 for predicted noise) with 0.004", 0.009", 0.025", and 0.05" spatial scales. Figures 4 and 5 show predicted signal-to-noise ratio per wavelength channel (R~4000) for a given aperture size in 5 hours of integration for four broadband observations in the finest (0.004") and coarsest (0.05") spatial scales, respectively. The 0.004" spatial scale uses a different aperture size for each broadband ($2\lambda_{cen}/D$). The PSF at coarser scales (0.025" and 0.05") is under-sampled and a fixed aperture size of 0.1" was chosen for all four broadbands. It is important to note that the under-sampled PSF makes the signal-to-noise ratio sensitive to pixel-to-pixel boundaries. Since the AO diffraction core is much smaller than a pixel, if it falls on a pixel center, it competes against 2 times less noise than if it falls off center.

The background is assumed to be the average between OH-sky emission lines for all four filters. The combination of higher Strehl ratios at J and H broadband compared to Y broadband, and being able to sample spectral features between the OH-sky emission allows for higher signal-to-noise per wavelength to be achieved at J and H filters. The effect of thermal background at K broadband becomes particularly significant at coarser scales since the sky coverage is ~15 times greater than the finest spatial scales. At shorter wavelengths (< 1μm), the signal-to-noise decreases since the background between OH-sky emission is dominated by zodiacal emission and the Strehl ratio decrease by a factor of 2.

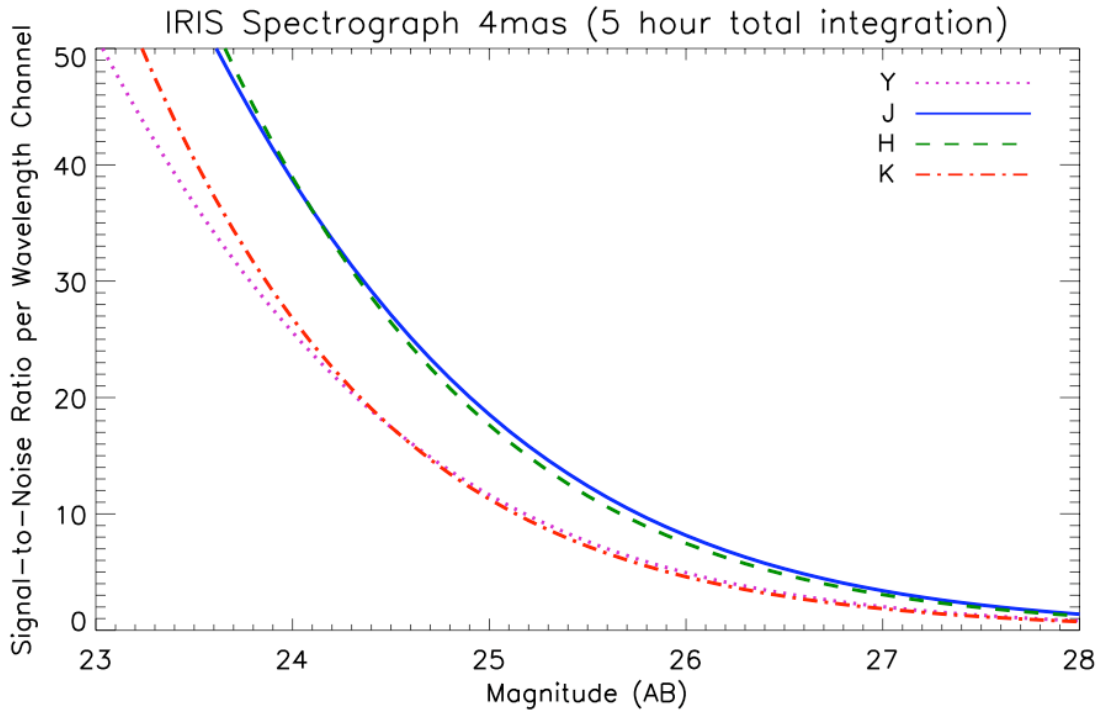

**Figure 4**: Estimated signal-to-noise ratio per wavelength channel versus a given magnitude (AB) illustrating the sensitivities for the integral field spectrograph at the 0.004" per spatial element scale in each broadband filter (Y, J, H, K) using a 2λ/D aperture size over a single point source. The central wavelength of the broadband filter is used to define the aperture size. A total integration time of 5 hours was made up of single exposures of 900 seconds stacked 20 times in Y, J, and H, and K.

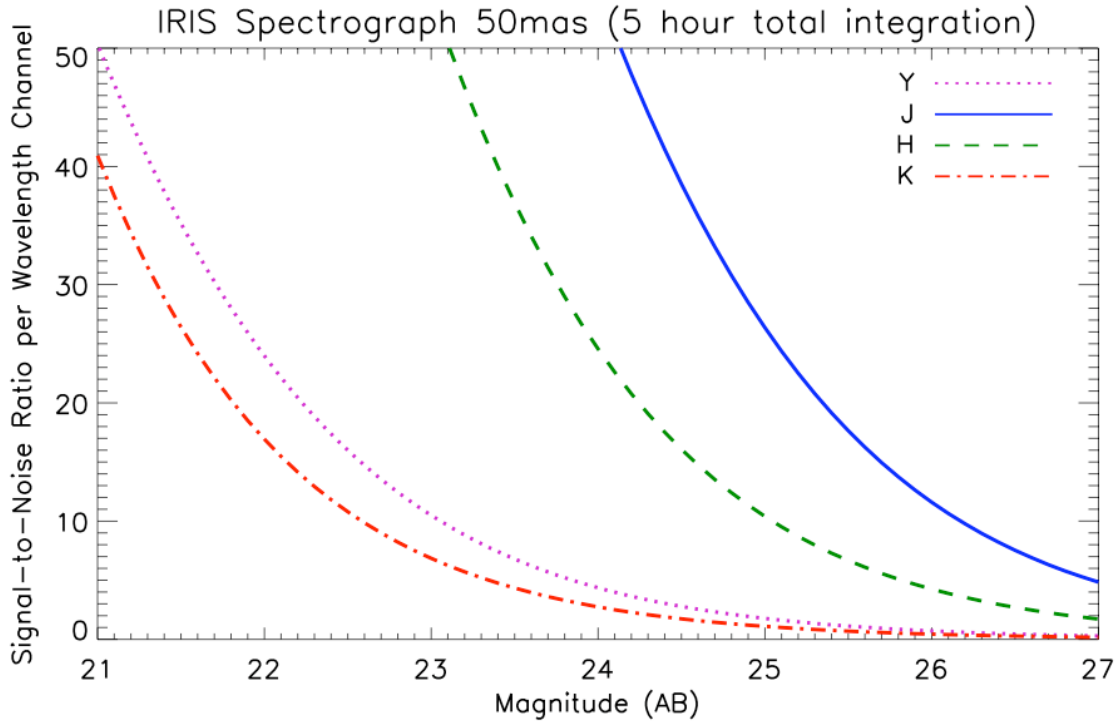

**Figure 5**: Estimated signal-to-noise ratio per wavelength channel versus a given magnitude (AB) illustrating the sensitivities for the integral field spectrograph at the 0.05" per spatial element scale in each broadband filter (Y, J, H, K) using a fixed aperture size of 0.1" over a single point source. A total integration time of 5 hours was made up of single exposures of 900 seconds stacked 20 times in Y, J, and H, and K. The coarser scale yields a higher background compared to the finer spatial scales and undersamples the PSF, therefore decreasing the signal-to-noise in these simulations.

## 5. SPECTROGRAPH RESOLVED SOURCE SENSITIVITIES

Exploring the performance of IRIS on resolved sources is not trivial since there are innumerable combinations of flux distributions and surface brightness profiles for any given astronomical object. We have developed an IRIS 3D data simulator to investigate different science cases and sensitivities for resolved sources. The simulator includes background, noise, throughput of the integrated system, blaze function, and PSF as a function of wavelength (as described in Section 2).

An important consideration for IFS sensitivities on resolved sources is the surface brightness profile and the number of spaxels over which the source is distributed. For instance, a half-arcsecond (0.5") resolved object (typical size of high-redshift galaxies) at the finest spatial scales (e.g., 0.004") would be sampled by 15,625 spatial elements compared to 100 spaxels using the coarsest scale (0.05"). If sources have cuspy or peaky surface brightness distributions as opposed to smooth profiles, this increases the signal-to-noise achieved per spatial element. In this section, we investigate these effects using simulated IRIS observations of high-redshift galaxies with a range of star formation rates, velocity dispersions, and surface brightness profiles.

### 5.1 High-Redshift Star Forming Galaxies (1 < z < 5)

With the recent advent of near-infrared integral field spectroscopy there have been numerous research programs to study the 2D dynamics and chemical abundance history of high-redshift (1 < z < 3) star forming galaxies using rest-frame optical emission lines[10,11,12]. Such programs offer insight into a number of fundamental questions about the formation and evolution of early galaxies[9,13,14]. Star forming galaxies currently being studied on 8-10m telescopes typically have total integrated star formation rates of 10-100 M☉ yr$^{-1}$, stellar masses of $10^{9-11}$ M☉, and dark matter halo masses of $10^{11-13}$ M☉[15]. IRIS will offer the ability to probe to even lower star formation rates and passive "red" galaxies at higher redshifts ($1 < z < 5$).

To investigate what star formation rates will be observable with IRIS fine and coarse scales, we simulate Hα emission from z=1.5 (look back time 9.3 Gyr) galaxies with various integrated star formation rates over four spatial areas. A Kennicutt law is used to estimate the total luminosity of each source for a given integrated, dereddened star formation rate[16]. We then derive the observed flux using concordance cosmology[17]. Figures 6 and 7 show the signal-to-noise per spaxel per FWHM versus star formation rate for z=1.5 galaxies with a fixed velocity dispersion of 80 km s$^{-1}$, observed in H-band at the 0.004" and 0.05" spatial scales for 5 hours. Hα emission is redshifted into a region between OH-sky emission lines. Even with the finest scales, IRIS will be a powerful tool for studying luminous high-redshift star forming galaxies with star formation rates of 10-100 M☉ yr$^{-1}$. The 0.004" and 0.009" scales will offer the ability to observe galaxies that are currently being studied with 8-10m telescopes, but with an unprecedented spatial resolution of <100 parsecs.

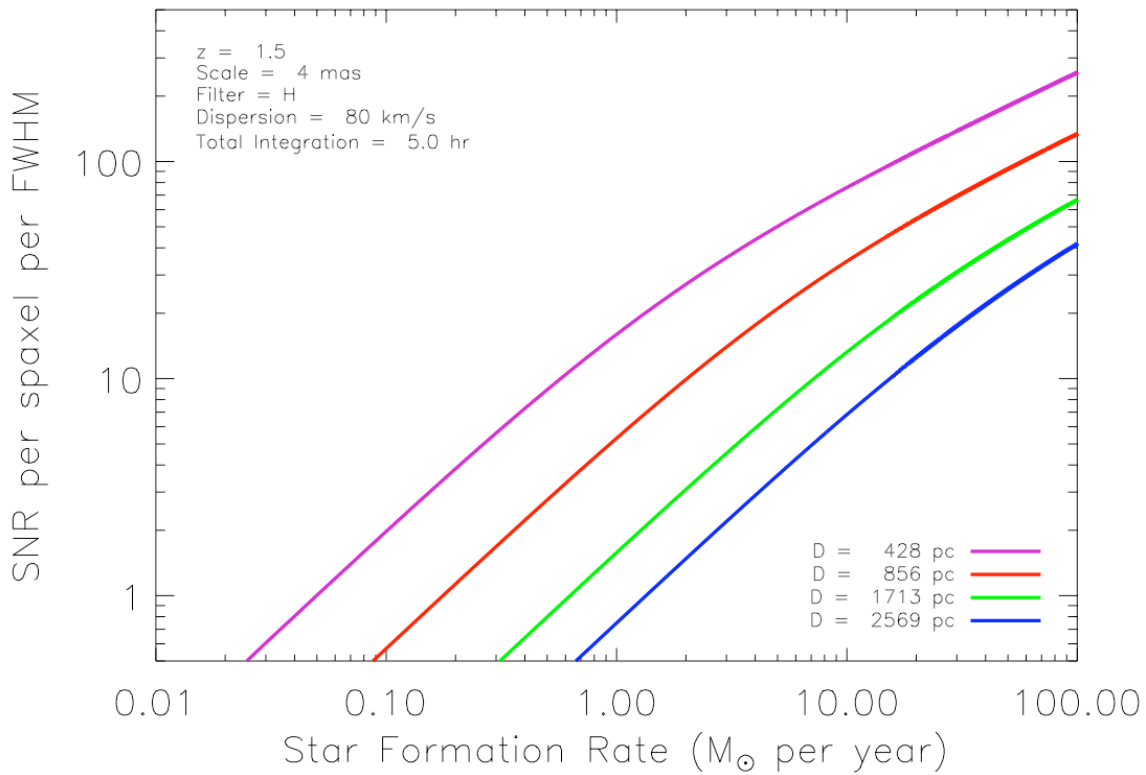

**Figure 6**: Estimated signal-to-noise ratio per spaxel per FWHM versus star formation rate for Hα emission in the 0.004" integral field spectrograph scale of a z=1.5 (look back time 9.3 Gyr) galaxy given a range of flux distribution sizes (angular diameters 0.05", 0.1", 0.2", and 0.3") and a fixed intrinsic velocity dispersion (80 km s$^{-1}$). A total integration time of 5 hours was made up of single exposures of 900 seconds stacked 20 times in H broadband. At the 0.004" spatial scale IRIS will be able to observe z=1.5 galaxies with star formation rates of 5 – 100 M yr$^{-1}$ with sufficient signal-to-noise.

IRIS coarse spatial scales will be powerful for studying galaxies with low and moderate star formation rates and at probing diffuse, faint regions and the outer extents of more massive high-redshift galaxies. This is illustrated in Figure 7, where the 0.05" scale will allow observations of z=1.5 star forming galaxies with integrated star formation rates of 0.1 – 10 M☉ yr$^{-1}$. Observing portions of galaxies with star formation rates below 1 M☉ yr$^{-1}$ will be achievable if the integrated star formation is confined within a 1kpc region. The 0.025" and 0.05" will also be advantageous for detecting fainter emission lines (e.g., [OII], Hβ, [OIII], [NII]) to study the 2D metallicity and dynamics of high-redshift galaxies.

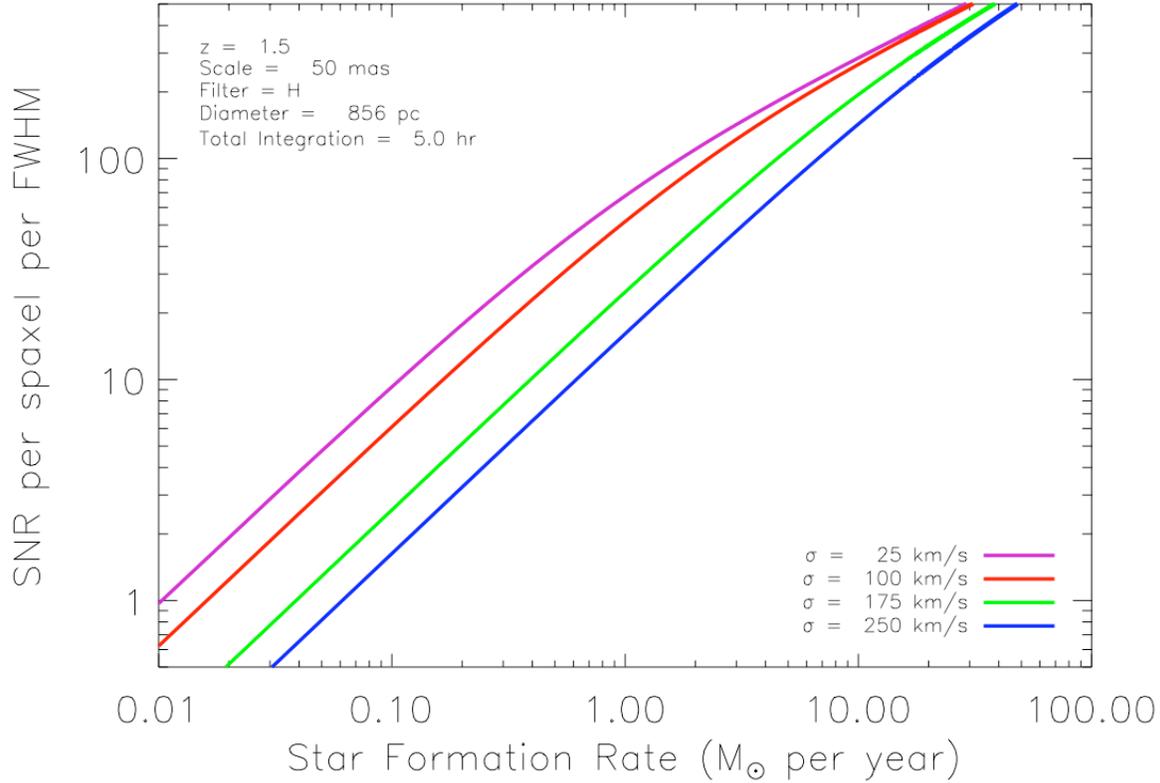

**Figure 7**: Estimated signal-to-noise ratio per spaxel per FWHM in the 0.05" integral field spectrograph scale for Hα emission of a z=1.5 (look back time 9.3 Gyr) galaxy versus star formation rate (M yr$^{-1}$) for a range of flux distribution sizes (angular diameters 0.05", 0.1", 0.2", and 0.3") and a fixed intrinsic velocity dispersion (80 km s$^{-1}$). A total integration time of 5 hours was made up of single exposures of 900 seconds stacked 20 times in H. At 0.05" spatial scales IRIS will be able to observe z=1.5 galaxies with star formation rates of less than 1 M yr$^{-1}$ with sufficient signal-to-noise.

The intrinsic velocity dispersion of a galaxy can also affect the sensitivity to individual galaxies, wherein larger velocity dispersions lower the overall signal-to-noise ratio. This is illustrated in Figure 8, where four different velocity dispersions (25, 100, 175, 250 km s$^{-1}$) for a z=1.5 galaxy yield a factor of 2-3 times greater signal-to-noise for a velocity dispersion that is 10 times smaller. For galaxies with star formation rates greater than 1 M☉ yr$^{-1}$, a velocity dispersion that is less than 300 km s$^{-1}$ will not significantly effect the signal-to-noise achieved.

IRIS will be able to extend 2D dynamical and chemical abundance studies out to even higher redshifts than are currently obtainable with integral field spectrographs on 8-10m telescopes. Figure 9, 10, 11, 12 present simulated Hα or [OII] (3727Å) emission from high-redshift (1< z < 5) galaxies observed using the 0.05" spatial scale. These galaxies templates were selected from local compact UV–luminous galaxies that were observed with Advanced Camera for Surveys on the Hubble Space Telescope with an Hα narrowband filter[18]. Each simulated galaxy observation consists of 20 individual 900 second exposures, for a total of 5 hours of integration time. The signal-to-noise at each spaxel is calculated by collapsing over the spectral extent of the emission line. IRIS will be able to study star formation rates less than 0.01 M☉ yr$^{-1}$ per spaxel within the outer extents of z=1.0 and 1.5 galaxies. For example, IRIS is able to detect the secondary diffuse source ~20 kpc from the primary galaxy at z=1.0 (see Figure 9). At z=2.5, Hα is redshifted into the K band where the thermal background is rising and the signal-to-noise is more sensitive to the surface brightness profiles. IRIS coarsest scales will be able to reach galaxies with moderate star formation rates with sufficient signal-to-noise. Figure 11 shows the star formation rates reached (~0.05 M☉ yr$^{-1}$ per spaxel) on z=2.5 galaxy in a 5 hour total integration. IRIS will, for the first time, extend spatially-resolved observations to even higher redshifts (z > 4) with adequate signal-to-noise for dynamical studies. Figure 12 presents the signal-to-noise achieved on [OII] emission redshifted into the K-band for a z=5 galaxy with an integrated star formation rate of 30 M☉ yr$^{-1}$. Though these integrated star formation rates are much higher than those that can be probed at lower redshift, the ability to study the dynamics at such high redshifts demonstrates the power of IRIS to answer fundamental questions about the formation and evolution of galaxies.

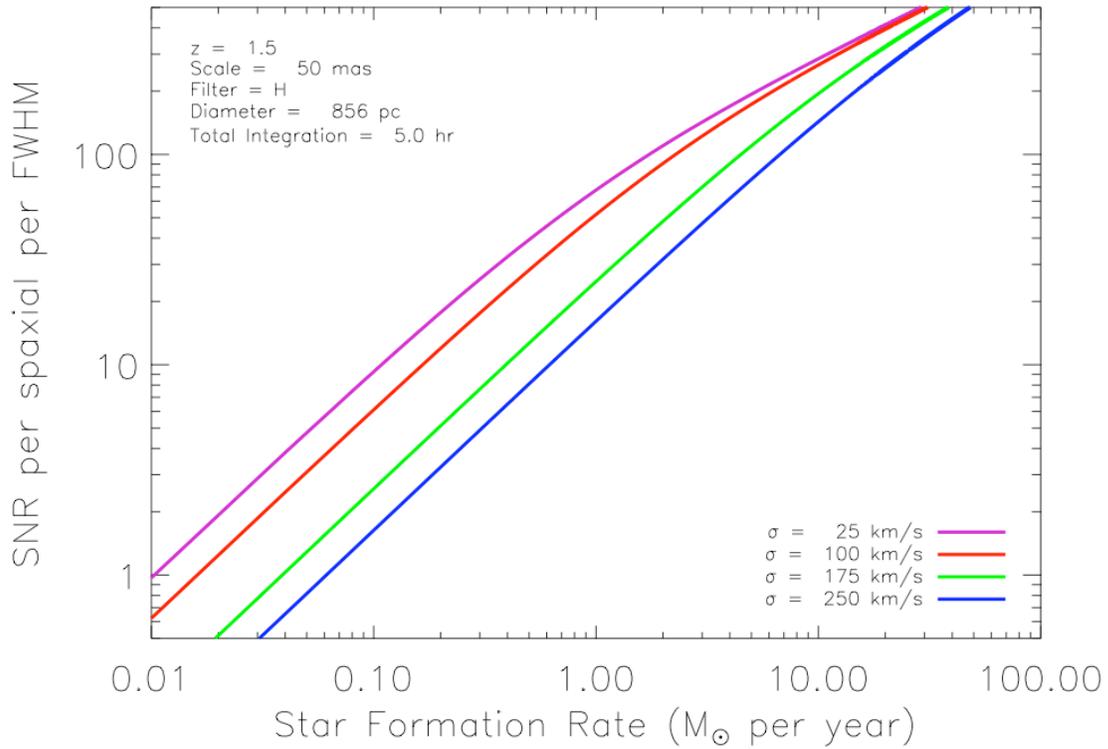

**Figure 8**: Estimated signal-to-noise ratio per spaxel per FWHM in the 0.05" integral field spectrograph scale for Hα emission of a z=1.5 (look back time 9.3 Gyr) galaxy given a range of intrinsic velocity dispersions (25, 100, 175, and 250 km s$^{-1}$) and a fixed surface brightness size of 0.1". Total integration time of 5 hours was made up of single exposures of 900 seconds stacked 20 times in H.

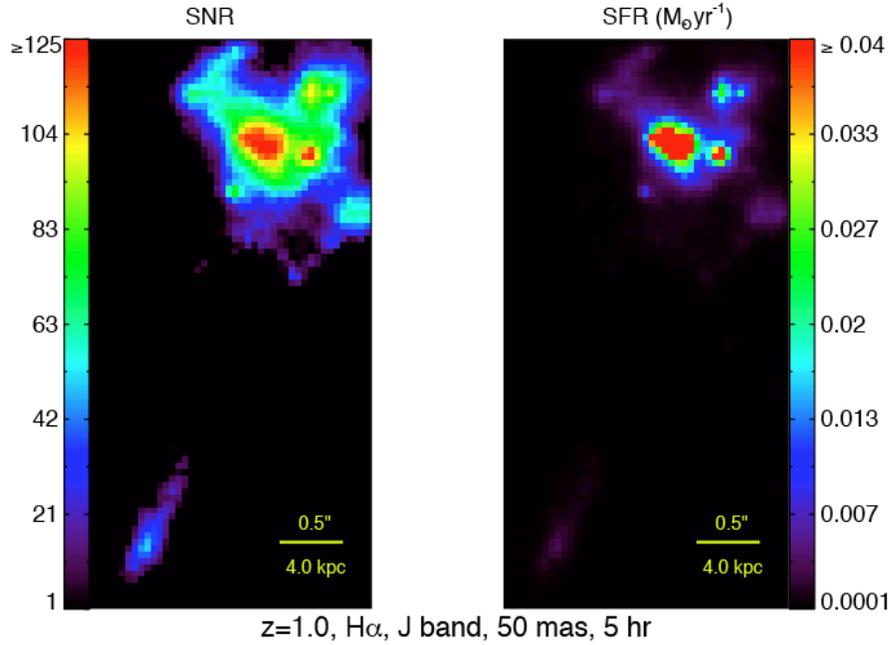

**Figure 9**: (Left) Integrated signal-to-noise ratio (SNR) for a z=1.0 (look-back time 7.7 Gyr) simulated galaxy's Hα emission redshifted into J band with an integrated star formation rate of 5 M yr$^{-1}$ and an intrinsic velocity dispersion of 80 km s$^{-1}$. Total integration time is 5 hours, with 20 stacked 900 second exposures. Note the diffuse region and the secondary object ~20 kpc away from the primary galaxy (blue-purple) that can be detected using IRIS in the coarsest scale (0.05"). (Right) Star Formation Rate (SFR; M⊙ yr$^{-1}$) spatially distributed across the galaxy.

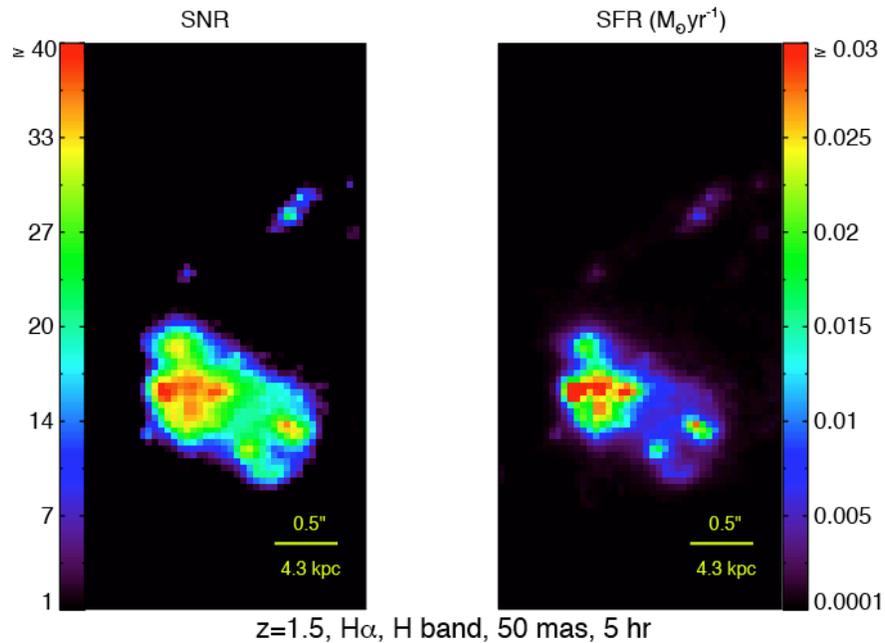

**Figure 10**: (Left) Integrated signal-to-noise ratio for a z=1.5 (look-back time 9.3 Gyr) simulated galaxy's Hα emission redshifted into H band with an integrated star formation rate of 5 M yr$^{-1}$ and intrinsic velocity dispersion of 80 km s$^{-1}$. Total integration time is 5 hours, with 20 stacked 900 second exposures. (Right) Star Formation Rate (SFR; M⊙ yr$^{-1}$) spatially distributed across the galaxy.

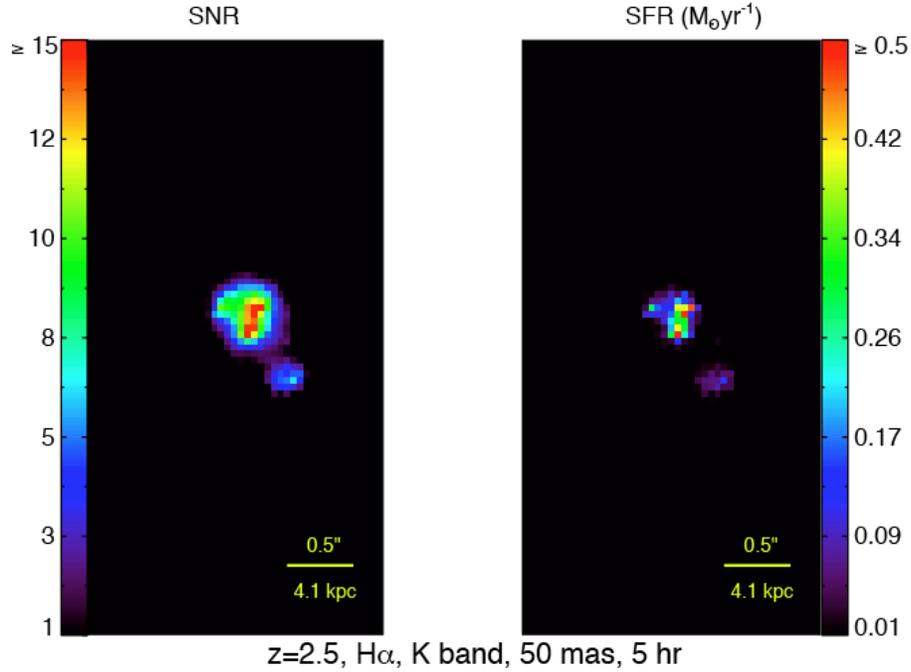

**Figure 11**: (Left) Integrated signal-to-noise ratio for a z=2.5 (look-back time 11 Gyr) simulated galaxy's Hα emission redshifted into K band with an integrated star formation rate of 10 M yr$^{-1}$ and intrinsic velocity dispersion of 80 km s$^{-1}$. Total integration time is 5 hours, with 20 stacked 900 second exposures. (Right) Star Formation Rate (SFR; M☉ yr$^{-1}$) spatially distributed across the galaxy.

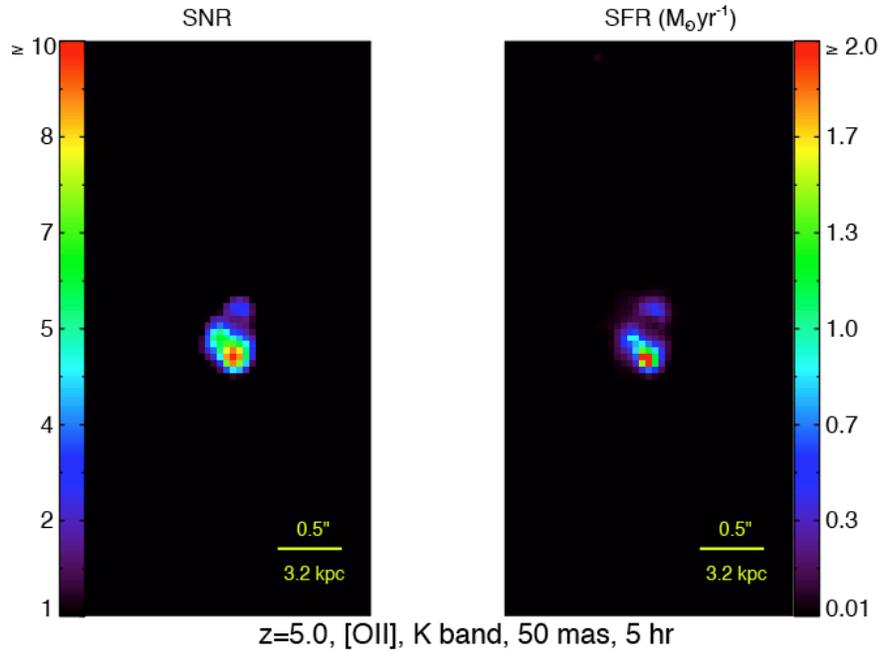

**Figure 12**: (Left) Integrated signal-to-noise ratio for a z=5.0 (look-back time 12.5 Gyr) simulated galaxy's [OII] emission redshifted into K band with an integrated star formation rate of 30 M yr$^{-1}$ and intrinsic velocity dispersion of 80 km s$^{-1}$. Total integration time is 5 hours, with 20 stacked 900 second exposures. (Right) Star Formation Rate (SFR; M☉ yr$^{-1}$) spatially distributed across the galaxy.

## 5.2 First Light Galaxies (7 < z < 12)

One of the primary science goals of IRIS is to observe the first galaxies and stars that formed from pristine primordial gas. Currently, studying these elusive objects has only been explored theoretically because of the lack of sensitivity of 8-10m class telescopes and current space-based facilities. But both TMT and JWST instrument capabilities will allow their detection to be viable[9,19]. JWST will offer an efficient means for surveying these sources, while TMT will have the advantage of using the diffraction-limit of a 30m telescope with fine spatial resolutions and high Strehl ratios to investigate morphologies and velocities of these early galaxies. Both Lyα and HeII (1640Å) emission from these early galaxies (z > 8; look back time < 700 Myr) are fortuitously redshifted into the near-infrared. The luminosity function and predicted flux distributions of these galaxies are highly uncertain. But using current theoretical predictions, both the IRIS imager and integral field spectrograph should be able to detect these sources in some of the prime near-infrared atmospheric and background windows. For instance, for a z=10 galaxy, using the 0.05" spatial scale in J-band a Lyα flux of $1 \times 10^{-18}$ erg s$^{-1}$ cm$^{-2}$ within ~400 pc (0.1") diameter will have a signal-to-noise per spaxel per FWHM (assuming σ=30 km s$^{-1}$) of 10 in a 5 hour total integration. When combining diffraction-limited near-infrared imaging and spectroscopy, IRIS on TMT will be a uniquely powerful instrument to study the first stars and galaxies in formation.

## ACKNOWLEDGEMENTS

The authors gratefully acknowledge the support of the TMT partner institutions. They are the Association of Canadian Universities for Research in Astronomy (ACURA), the California Institute of Technology and the University of California. This work was supported as well by the Gordon and Betty Moore Foundation, the Canada Foundation for Innovation, the Ontario Ministry of Research and Innovation, the National Research Council of Canada, Natural Sciences and Engineering Research Council of Canada, the British Columbia Knowledge Development Fund, the Association of Universities for Research in Astronomy (AURA), the U.S. National Science Foundation, and National Astronomical Observatory of Japan (NAOJ). The TMT project is planning to build the telescope facilities on Mauna Kea, Hawaii. The authors wish to recognize the significant cultural role and reverence that the summit of Mauna Kea has always had with the indigenous Hawaiian community. We are most fortunate to have the opportunity to conduct research from this "heiau" mountain.

## REFERENCES

[1] Larkin, J. et al., "OSIRIS: a diffraction limited integral field spectrograph for Keck", Proc. SPIE 6269, 62691A (2006)
[2] Eisenhauer, F. et al., "SINFONI - Integral field spectroscopy at 50 milli-arcsecond resolution with the ESO VLT", Proc. SPIE 4841, 1548 (2003)
[3] McGregor, P. et al., "Gemini near-infrared integral field spectrograph (NIFS)", Proc. SPIE 4841, 1581 (2003)
[4] Larkin, J. et al., "The Infrared Imaging Spectrograph (IRIS) for TMT: Instrument Overview", Proc. SPIE, 7735, (2010)
[5] Nelson, J. and Sanders, G. H., "The status of the Thirty Meter Telescope project", Proc. SPIE 7012, 70121A (2008)
[6] Suzuki, R. et al., "The Infrared Imaging Spectrograph (IRIS) for TMT: Imager Design", Proc. SPIE 7735, (2010)
[7] Herriot, G. et al., "NFIRAOS: TMT narrow field near-infrared facility adaptive optics", Adaptive Optics for Extremely Large Telescopes, (2010)
[8] Moore, A. et al., "The Infrared Imaging Spectrograph (IRIS) for TMT: Spectrograph Design", Proc. SPIE 7735, (2010)
[9] Barton, E. et al., "The Infrared Imaging Spectrograph (IRIS) for TMT: The Science Case", Proc. SPIE 7735, (2010)
[10] Förster-Schreiber, N., et al., "The SINS Survey: SINFONI Integral Field Spectroscopy of z ~ 2 Star-forming Galaxies" ApJ 706, 1364 (2009)

[11] Law, D. R., et al., "The Kiloparsec-scale Kinematics of High-redshift Star-forming Galaxies", ApJ 697, 2057 (2009)
[12] Wright, S. A., et al., "Dynamics of Galactic Disks and Mergers at z ~ 1.6: Spatially Resolved Spectroscopy with Keck Laser Guide Star Adaptive Optics", ApJ 699, 421 (2009)
[13] Law, D. R., et al., "Kinematics and Formation Mechanisms of High-Redshift Galaxies", The Astronomy and Astrophysical Decadal Survey, Science White paper no. 172 (2010)
[14] Wright, S. A., et al., "Tracing the Evolution and Distribution of Metallicity in the Early Universe", The Astronomy and Astrophysical Decadal Survey, Science White paper no. 327 (2010)
[15] Adelberger, K., et al., "The Spatial Clustering of Star-forming Galaxies at Redshifts 1.4 <~ z <~ 3.5", ApJ 619, 697 (2005)
[16] Kennicutt, R. C., "The Global Schmidt Law in Star-forming Galaxies", ApJ 498, 541 (1998)
[17] Komatsu, E., et al. "Seven-Year Wilkinson Microwave Anisotropy Probe (WMAP) Observations: Cosmological Interpretation", arXiv, 1001.4538 (2010)
[18] Overzier, R., et al., "Hubble Space Telescope Morphologies of Local Lyman Break Galaxy Analogs. I. Evidence for Starbursts Triggered by Merging", ApJ 677, 37 (2008)
[19] Cooke, J., et al., "First Light Sources at the End of the Dark Ages: Direct Observations of Population III Stars, Proto-Galaxies, and Supernovae During the Reionization Epoch", The Astronomy and Astrophysical Decadal Survey, Science White paper no. 53 (2010)